# The Interplay between Curvature Effect and Atomic Vacancy Defects in the Electronic Properties of Semi-Metallic Carbon Nanotubes

Hui Zeng, [†,‡,♀] Huifang Hu,[¶] and Jean-Pierre Leburton[*,‡]

[†]*College of Material Science and Engineering, Hunan University, Changsha 410082, China*

[‡]*Beckman Institute, University of Illinois at Urbana-Champaign, Urbana, Illinois 61801, USA*

[¶]*College of Physics and Microelectronic Science, Hunan University, Changsha 410082, China*

[♀]*Permanent address: College of Physical Science and Technology, Yangtze University, Jingzhou, 434023, China*

[*]Corresponding Email:jleburto@illinois.edu

**Abstract:** We investigate the electronic properties of semi-metallic (12,0) carbon nanotubes in the presence of a variety of mono-, di- and hexa-vacancy defects, by using first principle DFT combined with non-equilibrium Green's function technique. We show that defect states related to the vacancies hybridize with the extended states of the nanotubes to modify the band edge, and change the energy gap, resulting from the curvature effect. As a consequence, the nanotube conductance is not a monotonic function of the defect size and geometry. Paradoxically, hexa-vacancy nanotubes have higher conductance than di-vacancy nanotubes, which is due to the presence of mid-gap states originating from the defect, thereby enhancing the conductance.

**Introduction**

The unique electronic, mechanical, and transport properties of single walled carbon nanotubes [1] (*SWNTs*) make them promising candidates for the ultimate miniaturization of electronic functions at the molecular level [2–4]. Extensive experimental [1] and theoretical effects [5] have been deployed to unveil novel transport properties [6]. In particular, the ability to modify the nanotube properties through deformation, doping, or the creation of single- and multiatom vacancies has attracted considerable attention [6]. In this context it has been shown that electron [7] and ion irradiation [8] can be used to artificially remove individual carbon atoms from *SWNT*s. Recently, atomically re-solved Scanning Tunneling Microscopy (*STM*) has clearly shown that small holes can be created in metallic multiwall carbon nanotubes with diameter of about 10*nm* [9]. While the presence of carbon vacancies governs both the electronic [10] and mechanical properties [11] of nanotubes, they can be used to control the operation as chemical sensors [12]. Nanotube with vacancies also can be used as catalysts for thermal dissociation of water [13], whereas reconstruction of vacancies due to dangling bond saturation can provide active sites for atomic adsorption [14].

The structure and transformation energy of atomic vacancies in nanotubes, and especially, the existence of vacancy clusters are of fundamental importance for understanding the formation as well as the emergence and conditions for fractures in nanotubes [15]. Although some work has been devoted to carbon nanotubes with monovacancy [16] and divacancies, by contrast electronic properties of nanotubes with clusters of vacancies have received little attention [17]. Of special interest is the influence of atomic vacancy defects on the small band gap opened by the curvature of metallic(m,0) CNT's. Duo to their chirality, (m,m) CNT do not have this effect [6]. In this paper, we investigate by *ab initio* simulation, the combined influences of curvature effect and presence of vacancy defects of various sizes and geometries on the electronic properties of (12,0) semi-metallic CNTs.

Structure stability is assessed based on the length of the new bonds formed during reconstruction around the defects, and the nature of the defect states are studied by identifying and analyzing the band structure. Our results show that the current in nanostructures with six atoms removed and displaying a highly symmetric pattern is larger than SWNT with divacancy defects. While the point defect divacancy is the most stable because of its smallest transformation energy in contrast to the hexvacancy with largest transformation energy, that doesn't preclude that hexvacancies are energetically unfavorable in multi-walled nanotube. Higher-order defects can still be formed by removing a group of atoms at once with high energy impacts or chemical etching [15,18].

**Model**

Our study focuses on zigzag (12,0) SWNT's with 10 Å diameter, using a supercell L = 42.6Å (10 unit cells). The optimization structure calculations are performed in the framework of density functional theory, using a basis set of numerical pseudoatomic orbitals as implemented in the SIESTA code [19]. Standard norm-conserving Troullier-Martins [20] pseudopotentials orbitals are used to calculate the ion-electron interaction. The exchange correction energy is calculated within the generalized gradient approximation, as parameterized by Perdew et al [21]. We adopt a double-z polarized basis set with an energy cutoff (for real space mesh points) of 200Ry for structural relaxations. To test the accuracy of the structure of defected tubes after reconstruction, we compare our relaxed structures with published works and found good agreement [17,22].

The $I-V$ characteristics of the nanotube device are obtained for a two-probe device geometry where the central region contains the vacancy defects and both leads consist each of one supercell pristine tube (see Fig.1). The current is calculated by means of the Laudauer formula:

$$I = \frac{2e}{h}\int_{\mu_{min}}^{\mu_{max}} dE(f_l - f_r)T(E)$$

where the transmission coefficient T as a function of the electron energy E is obtained by using self-consistent real space, nonequilibrium Green's function formalism and the density functional theory [23].

$f_l$ ($f_r$) is the corresponding electron distribution function of the electron eigenstates of the left (right) electrode, and $\mu_{min}$=min($\mu$+V(b),$\mu$) [$\mu_{max}$=max($\mu$+V(b),$\mu$)] denotes the minimum (maximum) electrochemical potential $\mu$ of the electrodes.

**Results and discussion**

In Fig.2, our structure optimization calculations show different configurations of carbon vacancies i.e. monovacancies (1Va, 1Vb), divacancies (2Va, 2Vb, 2Vc), hex-vacancies (6Va, 6Vb) obtained by spontaneously reconstructing stable or metastable structures. Table 1 lists the calculated transformation energy [24] of all configurations. In the case of monovacancies (1Va and 1Vb), the reconstruction around the defect forms a dangling bond (*DB*), and leaves a so called 5−1*DB* defect, as obtained from our optimization procedure which is in agreement with existing data [25,26]. This dangling bond is available to form a bridge for connecting two tubes [14] or to provide an active site for atomistic adsorption. The 1Va configuration that only contains a 5−1*DB* defect is one of the most stable structures, compared with two Stone-Wales defects (pentagon-heptagon pair defects) in the 1Vb configuration. One can estimate the transformation barrier for the Stone-Wales defects to be 12.41*eV*−5.85*eV*=6.56*eV* [25]. In the case of divacancies (2Va,2Vb,2Vc), our simulation shows that except for the 2Vc configuration, four uncoordinated carbon atoms around the missing carbon defects have rebonded together to stabilize the nanostructure. Hence, in zigzag (12,0) nanotubes the 2Va configuration oriented perpendicular to the tube axis is energetically possible in addition to the tilted divacancies orientation, shown in 2Vb. The 2Vb not only contains a so-called 5−8−5 defect (pentagon-octagon-pentagon(Fig.2)), but also a Stone-Wales defect. It is interesting to note that the 2Vc configuration generates two pentagons around the defect as well as a carbon atom with only one bond to the neighboring atom. We believe this structure is chemical metastable since it provides a site for adsorbing other specimen atoms such as oxygen or nitrogen atoms. The 5−8−5 defects in divacancies are the most favorable structure in small diameter tubes [27]. For the hexa-vacancy cluster, the 6Va configuration contains two unsaturated atoms compared with 6Vb with six unsaturated atoms. As for the comparison between unrelaxed and relaxed structures, we find the area of defects become smaller after optimization, but the diameter and length of the simulated sample do not change at all. Therefore, it can be seen that the existence of six missing atoms only affects the local nanotube electronic structure. On the other hand, the 6Va structure manifests as a big hole with 14 number rings connected to two adjacent quadrangles, while the length of the whole tube shortens 0.1Å after relaxation. The diameter of the structure near the defect drastically shrinks which originates from the large hole. Consequently, the diameter of the tube at the sample edges changes from circular to elliptic.

**Table 1.** Transformation Energy

| Configurations | Transformation Energy (eV) |
| :---: | :---: |
| 1Va | 5.85 |
| 1Vb | 12.41 |
| 2Va | 3.69 |
| 2Vb | 9.15 |
| 2Vc | 10.72 |
| 6Va | 21.31 |
| 6Vb | 18.97 |

The charge density for each configuration is also shown in Figure.2 with color code for the charge distribution. Starting from the red to the blue in increasing densities. It is seen that the pristine tube as well as the 2Va, 6Va and 6Vb structures has a perfectly symmetric charge distribution. Specifically, it indicates that the charge density near the defect in all configurations is not very affected by the vacancy. More importantly, we find the charge density around the defects in 1Va and 6Vb configurations is much larger than that in divacancy configurations.

In Fig.3, the energy band structures of the 7 defective (12,0) tubes are displayed with the pristine (12,0) band structure for comparison. The dotted line in the figure indicates the Fermi level. In the pristine (12,0) tube, the highest occupied (lowest unoccupied) band is labeled $\alpha$ ($\alpha'$); both are doubly degenerate, and cross at the $\Gamma$-point, which contributes to two quantum transmission channels. In the 1Va structure, the defect lifts the degeneracy of the two doubled degenerate $\alpha$ and $\alpha'$ electron bands at the $\Gamma$-point, which are now labeled $\beta$ and $\beta'$ bands. Concomitantly, the defect gives rise to a new band labeled $\gamma$, the lowest unoccupied band that crosses the $\beta$ band at the $\Gamma$-point [28]. This $\gamma$-band results from quasi bound unsaturated s orbitals [16], and is partially occupied near the $\Gamma$-point at the expense of the $\beta'$-bands. The band structure of the 1Vb CNT is far more complex, especially above the Fermi level, where due to the reconstruction and the large size of the Stone-Wales defect, two new bands emerge. The flat energy band close to the Fermi level is called $\delta$ and opens the band gap by anti-crossing with the upper $\beta$-band a little away from the $\Gamma$-point. We label the next higher band as $\gamma$ and $\gamma'$ that anti-cross in the middle of the Brillouin zone, while the $\beta$-bands have moved to higher energy, as a result of significant hybridization among the original $\alpha$-bands and the defect states. In the 2Va nanotube, the band structure resembles the 1Va structure with a large splitting of the $\beta$-bands at the $\Gamma$-point. Here, the $\gamma$-band has disappeared, and the $\delta$-band is very close and below the Fermi level with a very small dispersion that flattens near the X-point; this flat $\delta$-state, which is also predicted by Berber et al [22] is partially occupied at the expense of the $\beta'$-bands near the $\Gamma$-point. The existence of Stone-Wales defects in the 2Vb configuration opens the band gap by shifting up the defect $\delta$-state and the $\beta'$-bands. Here, Stone-Wales defects produce only one $\gamma$-band located above the $\delta$-state. The band structure of the 2Vc configuration strongly resembles the 2Va band structure with the $\beta$ and $\beta'$ bands and the $\delta$-state, except for the existence of the band gap at the $\Gamma$-point in the former. Both 2Va and 2Vc don't show the $\gamma$-bands of the Stone-Wales di-vacancy defects. In the 6Va configuration, the splitting between the two $\beta'$-bands increases over the whole Brillouin zone; both the $\gamma$-band and the $\delta$-state reappear as in the 2Vb CNTs with the presence of a large gap. Here, however the $\beta$-bands below the Fermi level are quasi-degenerate. The 6Vb configuration consists of a large hole in the nanotube, which produces a new defect state labeled $\delta'$, very close to its parent d-state, above the Fermi level.

Fig.4 shows the density of states (*DOS*) on a finer scale around the gap for each vacancy configuration with the (12,0) pristine nanotube. Due to the *SWNT* curvature effect, the pristine semimetallic (12,0) tube has a small *DOS* gap of about $0.07eV$ [29], not shown on Fig.3 because of the larger scale. This small gap is symmetric with respect to the Fermi level, and the (*DOS*) exhibits two 1D singularities at the edge of $\alpha$ ($\alpha'$) bands. The split in the two 1D peaks is due to the lifting of degeneracy of the $\alpha$ ($\alpha'$) bands by the curvature effect. In the 1Va configuration the presence of the $5-1DB$ defect results in a loss of symmetry of the DOS around the Fermi level, and gives rise to two extremely high *DOS* peaks around the gap. However, the *DOS*-gap is smaller compared with the gap in the pristine tube, which is paradoxical given the presence of the defect. We attribute this band gap reduction to the presence of the $\gamma$-state that hybridizes with the $\beta$ and $\beta'$ bands. Similar effects have been recently predicted in graphene, where the presence of mid-gap states due to vacancy defects enhances the metallic character of the material [30]. In addition to the two previous peaks, another *DOS* peak is located at $-0.028eV$. In the 1Vb case, the gap increases again to $0.07eV$ due to the presence of the two Stone-Wales defects. We also observe two peaks located at about $0.032eV$ and $0.03eV$, below and above the Fermi level, respectively. Generally speaking, the main effect of the Stone-Wales defects in the 1Vb structure is to increase the gap compared to the 1Va, and substantially decrease the DOS around the band edges.

In the divacancy configurations, the 1D *DOS* singularities of the pristine structure around the gap broaden into shoulders. In both the 2Va and 2Vb structures, there is a *DOS* peak located at $-0.027eV$, which are caused by the presence of the 5−8−5 defects, while the second *DOS* peak located at $0.013eV$ in the 2Vb structure derives from the Stone-Wales defect. The symmetric 5−8−5 defects in the divacancy structures substantially lower the DOS that is still symmetric around the Fermi level with a gap of $0.066eV$. In the 2Vb configuration, the additional Stone-Wales defect results in the *DOS*-gap increasing to $0.078eV$, but it also loses its symmetry around the Fermi level. The defects in the 2Vc configuration also break the DOS symmetry around the Fermi level and give rise to two peaks at $0.029eV$ and $-0.032eV$, respectively, which are essentially due to the existing unbound C-atom. Here, the gap is unchanged compared to the 2Vb structure. For the three divacancy structures, there no dangling carbon atom, so no resonant scattering near the Fermi level as in the 1Va configuration is observed.

In the case of the 6Va structure, the gap is reduced to $0.042eV$, and is symmetric around the Fermi level, except for a large DOS peak at $0.018eV$ and a smaller peak at $-0.013eV$. Such a big defect induces a sizeable modification of the 6Va configuration around the radial direction of the tube, which strongly affects and substantially reduces the DOS near the Fermi level. In the 6Vb configuration, the gap is reduced to $0.058eV$, whereas the dip in the DOS at $0.034eV$ originates from resonant scattering by quasibound states [31]. Note that the presence of the two 1D *DOS* singularities in the valence band is practically unaffected by the defect, but lightly more split than in the pristine nanotube.

Fig.5 displays the $I-V$ characteristics of the (12,0) structures over a voltage range of $6mV$, which corresponds to an electric field range of $15KV/cm$ ($1meV$ equal to $2.5KV=cm$). All the $I-V$ characteristics are linear with different conductances (inset). As expected, the conductance of the pristine nanotube is the highest, closely followed by the 1Va conductance, which indicates that in addition to the reduced band gap (see Fig.4) the 5−1*DB* defect does not remarkably scatter electrons. On the opposite, the lower conductance of the 1Vb nanostructure is mainly due to the two Stone-Wales defects and its strong perturbation on the band structure of the pristine CNT. Unexpectedly, the 2Vb and 2Vc nanotubes exhibit the lowest conductance among all structures, while the 2Va configuration with symmetric defects shows the highest current, which is due to the smaller gap among the divacancy nanotubes, as well as the nature of its defect (5−8−5) compared to the Stone-Wales defect and the $DB$ in the 2Vb and the 2Vc configurations, respectively. Hence, the conductance difference between the 2Va and 2Vb configuration could be used to identify Stone-Wales defects [32]. Quite generally, in both the monovacancy and divacancy structures, the presence of the Stone-Wales defect arising from the atomic reconstruction around the vacancy lowers the conductance compared to the 5−1$DB$ and the 5−8−5 defect structures, which is in agreement with previous work [33,34].

However, the unexpected feature of our work is the superior conductance of the hexa-vacancy configurations compared with the 1Vb and divacancy structures, which is due to the geometry of the defect with highly symmetric patterns (Fig.2), which affects the band and DOS as the larger number of missing atoms reduces the curvature effects. Hence the 6Va nanostructure has a gap as small as in the 1Va CNT (Fig.4), which boosts the conductance to the level comparable to the 2Va CNTs. In the 6Vb structure the effect of the larger gap is offset by the presence of the $\delta$-$\delta'$ bands (Fig.3) that enhances conductance, especially in the valence band.

**Conclusion**

In conclusion, we have shown the conductance variation due to the presence of atomic vacancies in carbon nanotubes is not monotonous function of the number of missing C-atoms, but of the reconstruction around the defect and its spatial symmetry. Therefore, the ability to tailor the atomic structure of carbon nanotubes provides new ways to control their transport properties.

**Acknowledgement**

The authors thank Prof. A. Bezryadin, Dr. J. W. Wei and T. Markussen for helpful

discussion. We also thank Z. Y. Wang and Marcelo A. Kuroda for technical assistance in the MAC OS X Turing cluster. This work is supported by National Basic Research Program Grant No. 2006CB921605. Hui Zeng also thanks China Scholarship Council for pursuing his study in the University of Illinois at Urbana-Champaign as a joint PhD candidate.

**Figures and Figure Captions**

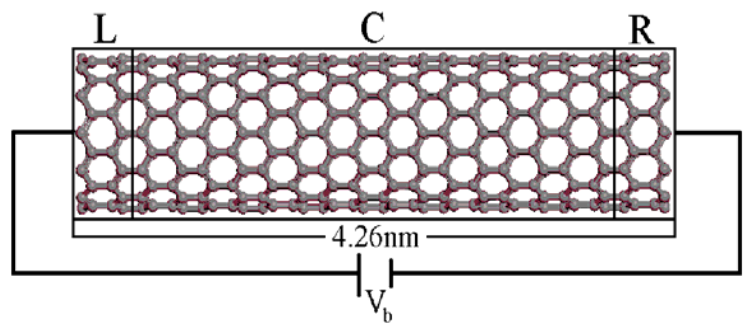

**Figure 1.** Schematic structure of the two-probe transport model, two semi-infinite left (L) and right (R) electrodes are comprised by one unit cell and extend to $z=\pm\infty$ with periodic boundary condition. The direction of transport is denoted by z.

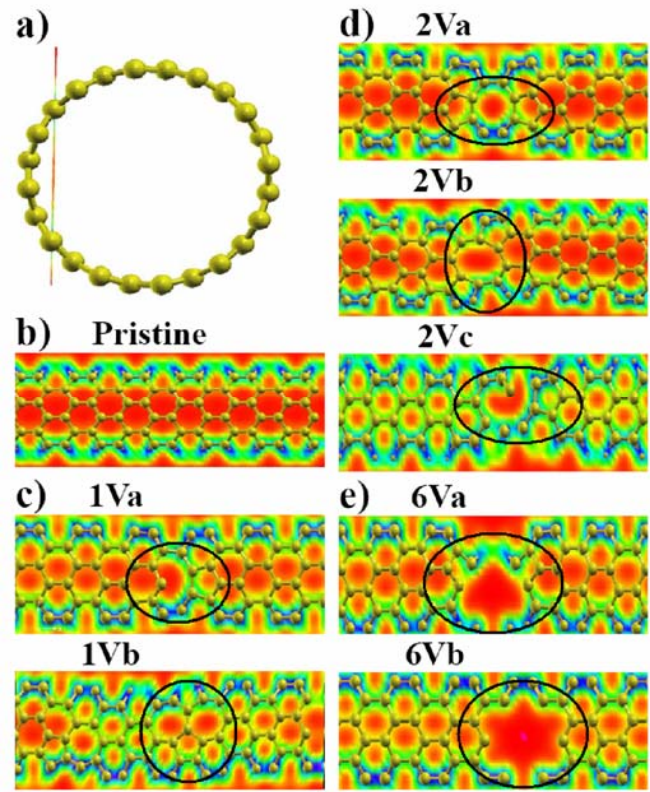

**Figure 2.** (Color online) Charge density of all structural configurations. (a) Ball-and-stick model for (12,0) SWNT. (b), (c), (d) and (e) are charge densities of (12,0) SWNT with no-vacancy, monovacancy, divacancy and hexa-vacancy, respectively. The defect areas are highlighted by the elliptical solid line. Color blue represents high density; the red low density.

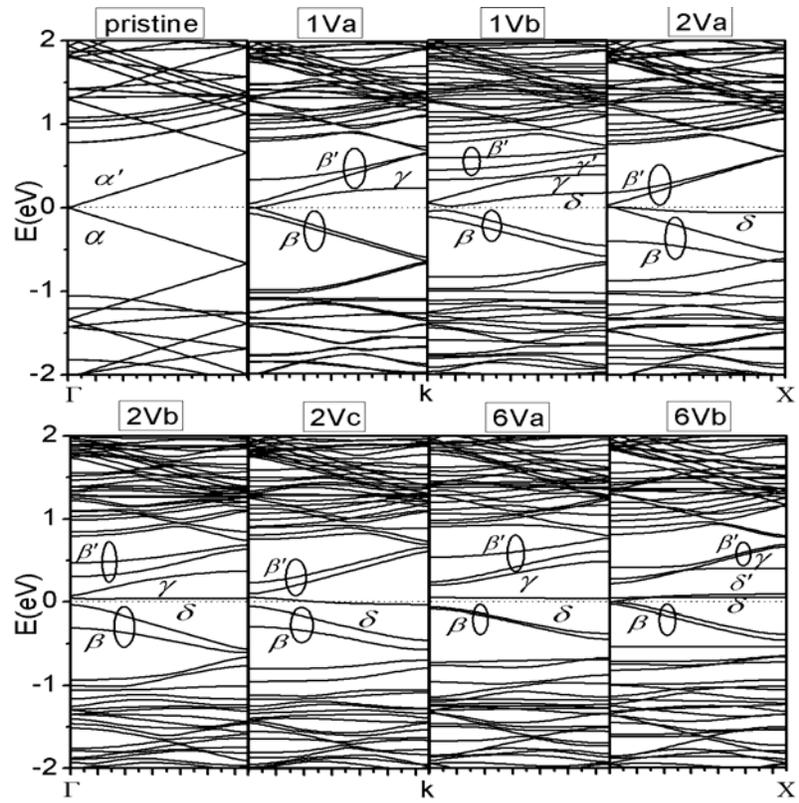

**Figure 3.** Band structures of (12,0) SWNTs with various vacancies. The band structure of the pristine (12,0) CNT is given for comparison.

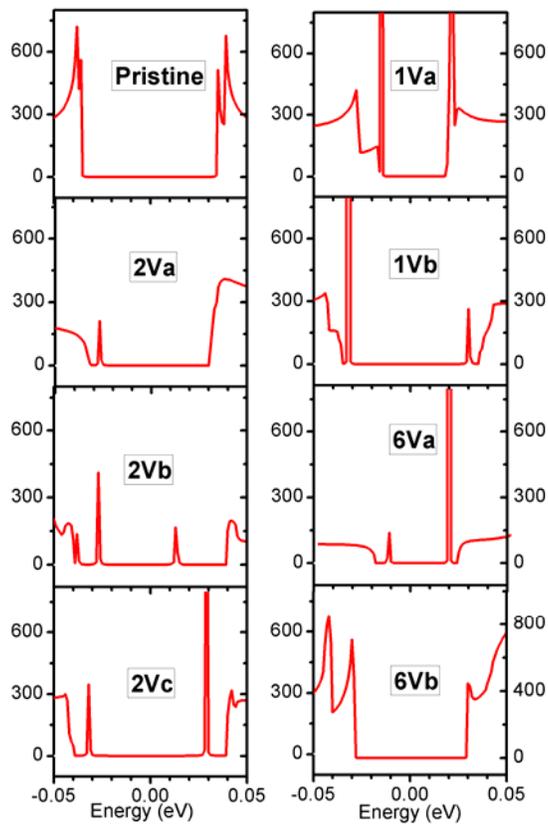

**Figure 4.** (Color online) DOS (Red solid lines) of (12,0) defective tubes as a function of energy, DOS of the pristine (12,0) CNT is given for comparison.

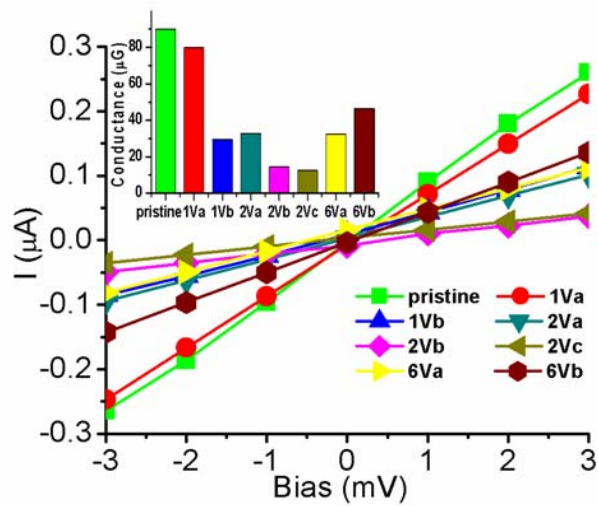

**Figure 5.** (Color online) I-V characteristics and conductance (Inset map) of (12,0) defective tubes; the results of pristine (12,0) CNT are given for comparison. The color in the I-V curve and conductance (Inset map) denotes the same structure.

15